\documentclass[12pt]{JHEP}
\usepackage{longtable}

\newcommand{\be}{\begin{equation}}
\newcommand{\ee}{\end{equation}}
\newcommand{\bea}{\begin{eqnarray}}
\newcommand{\eea}{\end{eqnarray}}

\newcommand{\nn}{\nonumber \\}

\newcommand{\beq}{\begin{equation}}
\newcommand{\eeq}{\end{equation}}
\newcommand{\beqa}{\begin{eqnarray}}
\newcommand{\eeqa}{\end{eqnarray}}
\newcommand{\beqan}{\begin{eqnarray*}}
\newcommand{\eeqan}{\end{eqnarray*}}
\newcommand{\ba}{\begin{array}}
\newcommand{\ea}{\end{array}}

\newcommand{\ve}{\varepsilon}
\newcommand{\vp}{\varphi}

\newcommand{\cL}{{\cal L}}

\newcommand{\no}{\nonumber}
\newcommand{\bdm}{\begin{displaymath}}
\newcommand{\edm}{\end{displaymath}}
\newcommand{\lgl}{\langle}
\newcommand{\rgl}{\rangle}

\newcommand{\dg}{\dagger}
\newcommand{\cE}{{\cal E}}

\newcommand{\ccdot}{\hskip-0.3ex\cdot\hskip-0.3ex}

\preprint{
LU TP 99-02\\
UWThPh-1999-02\\
ZU--TH 9/99\\
hep-ph/9902437\\
Revised April 1999}
\title{
The Mesonic Chiral Lagrangian of Order $p^6$
\thanks{Work supported in part by TMR, EC-Contract No. 
ERBFMRX-CT980169~(EURODA$\Phi$NE).}
}
\author{Johan Bijnens\\Dept. of Theor. Phys. 2, Lund University,\\
S\"olvegatan 14A, 
S-22362 Lund, Sweden}
\author{Gilberto Colangelo\\Inst. Theor. Physik, Univ. Z\"urich,
Winterthurerstr. 190,\\ 
CH-8057 Z\"urich-Irchel, Switzerland}
\author{Gerhard Ecker\\
Inst. Theor. Phys., Univ. Wien, Boltzmanng. 5,\\ A-1090 Wien,
Austria}

\abstract{We construct the effective
chiral Lagrangian for chiral perturbation theory
in the mesonic even-intrinsic-parity sector at order $p^6$. The
Lagrangian contains 112 in principle measurable + 3 contact terms 
for the general case
of $n$ light flavours, 90+4 for three and  53+4 for two flavours.
The equivalence between equations of motion and field redefinitions 
to remove spurious terms in the Lagrangians is shown to
all orders in  the chiral expansion.
We also discuss and implement other methods 
for reducing the number of terms to a minimal set.}


\begin{document}

\section{Introduction}\label{sec:intro} 
The low-energy limit of the theory of the strong interaction, QCD, is
chiral perturbation theory (CHPT). This is the effective field theory method
of solving in a long-distance expansion the Ward identities of the
chiral symmetry of QCD.
CHPT in the meson sector \cite{wein79,gl84,gl85}
\footnote{An overview of review articles and lectures as well as recent
results can be found in \cite{badhonnef}.}
is now being carried out at next-to-next-to-leading order. By now, 
several complete calculations to $O(p^6)$ exist \cite{p6su2,p6su3} and
the renormalization of the generating functional of Green functions of
quark currents will soon be available to this order \cite{bce3}. The
double divergences proportional to $1/(d-4)^2$ are already known \cite{bce4}.

In each specific calculation, the local contribution of $O(p^6)$ is a 
rather trivial part when compared to the two-loop contributions.
In those calculations, no attempt is usually made to
relate the low-energy constants of $O(p^6)$ in the local amplitudes
to those appearing in other processes. This is precisely the purpose
of the present paper, to establish the most general local solution of
chiral Ward identities at the level of the generating functional. This
solution amounts to constructing the effective chiral Lagrangian of 
$O(p^6)$, which is also invariant under Lorentz transformations, 
parity ($P$) and charge conjugation ($C$). We use here the external
field method of \cite{gl84,gl85} where only terms explicitly invariant
under these symmetries are relevant.

It is relatively easy to write down such chiral Lagrangians. The real
challenge is, of course, to find the minimal set of terms for those
Lagrangians. The corresponding low-energy constants then parametrize 
the most general local solutions of the chiral Ward identities. We
construct this Lagrangian first for a general number $n$ of light
flavours and then specialize to the phenomenologically relevant cases
$n=2$ and 3 where the number of independent terms is substantially
smaller. We confine ourselves to the Lagrangians of even intrinsic
parity, i.e., to terms without an $\ve$ tensor. We also compare our 
results with the work of Fearing and Scherer \cite{fs96} who have 
previously published chiral Lagrangians of $O(p^6)$ for general $n$ 
and for $n=3$.

In constructing the chiral Lagrangian of $O(p^6)$, we use 
partial integration (in the corresponding
action) and the equations of motion (EOM) of the lowest-order Lagrangian 
to reduce the number of chiral invariants. To make the large-$N_c$ 
counting transparent, where $N_c$ is the number of colours, we
employ the various relations among different
monomials of $O(p^6)$ to eliminate preferentially terms with multiple
flavour traces.\footnote{In the limit of large $N_c$,
terms with single flavour traces dominate. Each additional flavour trace
brings in a suppression of order $1/N_c$ \cite{largeNc}.}
The final Lagrangians are ordered essentially
according to the external fields in their components.

The paper is organized as follows. In Sect.~\ref{sec:sun}, we start 
by collecting 
the ingredients for constructing the effective chiral Lagrangian of 
$O(p^6)$ for chiral $SU(n)$. We use a basis where all (matrix)
operators have the same chiral transformation properties. In addition
to partial integration and the EOM, we use the Bianchi
identity for the field strength tensor on chiral coset space to
reduce the Lagrangian to a minimal form. We also extract the 
so-called contact terms that depend only on external fields. In
Sect.~\ref{sec:su23}, we simplify the Lagrangian further for chiral 
$SU(3)$ and
finally for $SU(2)$. It turns out that all 21 linear relations for
$n=3$ as well as the additional 37 relations for $n=2$ can be derived from
the respective Cayley-Hamilton relations. We discuss possible
applications and limitations of the chiral Lagrangians of $O(p^6)$ in
Sect.~\ref{sec:tour}. Sect.~\ref{sec:conc} contains some
conclusions. In App.~\ref{app:EOM} we present an explicit proof for
the equivalence between using the EOM and field transformations for
simplifying the Lagrangians. App.~\ref{app:CH} contains the complete
list of linear relations of the Cayley-Hamilton type for both $n=2$
and 3. Finally, we compare with the chiral Lagrangians of
Fearing and Scherer \cite{fs96} in App.~\ref{app:FS} and demonstrate
with some explicit examples why we end up with substantially fewer terms.

\section{Chiral $SU(n)$}\label{sec:sun}
In the formulation of Gasser and Leutwyler \cite{gl84,gl85}, the QCD
Lagrangian $\cL^0_{\rm QCD}$ with $n$ massless quarks is enlarged to
\begin{equation} 
\cL = \cL^0_{\rm QCD} + \overline q \gamma^\mu(v_\mu + a_\mu \gamma_5)q -
\overline q (s - ip \gamma_5)q \label{eq:QCD}
\end{equation} 
by coupling the quarks to external $n\times n$-dimensional
matrix fields in flavour space, $v_\mu,a_\mu,s,p$. 
At the quantum level, the theory exhibits a (local) chiral symmetry 
$SU(n)_L \times SU(n)_R\times U(1)_V$ that is spontaneously 
broken to $SU(n)_V\times U(1)_V $.

The basic building block of chiral Lagrangians is the Goldstone matrix
field $u(\vp)$ transforming as
\begin{equation} 
u(\vp) \to u(\vp')=g_R u(\vp) h(g,\vp)^{-1} 
= h(g,\vp) u(\vp) g_L^{-1} \label{eq:uphi}
\end{equation} 
under a general chiral rotation $g=(g_L,g_R) \in SU(n)_L \times SU(n)_R$ 
in terms of the compensator field $h(g,\vp)$. Mesonic chiral Lagrangians
can be constructed by taking (products of) traces of
products of chiral operators $X$ that either transform as
\begin{equation} 
X \to h(g,\vp) X h(g,\vp)^{-1} \label{eq:hXh}
\end{equation} 
or remain invariant under chiral transformations.
The simplest such operators are
\begin{eqnarray} 
\label{eq:ingr}
u_\mu &=& i \{ u^\dg(\partial_\mu - i r_\mu)u - 
u(\partial_\mu - i \ell_\mu) u^\dg\} \nn
\chi_\pm &=&  u^\dg \chi u^\dg \pm u \chi^\dg u  
\end{eqnarray}
with  $r_\mu = v_\mu + a_\mu$, $\ell_\mu = v_\mu - a_\mu$,
$\chi= 2 B (s+ip)$.
The mesonic chiral Lagrangian of lowest order can be written as
\begin{equation} 
 {\cal L}_2 = \frac{F^2}{4} \left\lgl u_\mu u^\mu+
 \chi_+\right\rgl \label{eq:L2}
\end{equation} 
with two low-energy parameters $F,B$ and with the usual notation
$\lgl \dots \rgl$ for the flavour trace.

In higher orders, we need additional operators for the construction of
chiral Lagrangians. Up to and including $O(p^6)$, the following
operators are sufficient for this purpose:
\begin{eqnarray} 
f_\pm^{\mu\nu} &=& u F_L^{\mu\nu} u^\dg \pm u^\dg F_R^{\mu\nu} u 
~, \qquad \nabla_\lambda f_\pm^{\mu\nu} \nn
h_{\mu\nu} &=& \nabla_\mu u_\nu + \nabla_\nu u_\mu \nn
\chi_{\pm \mu} &=&  u^\dg D_\mu \chi u^\dg \pm u D_\mu \chi^\dg u =
\nabla_\mu \chi_\pm -{i \over 2} \{\chi_\mp,u_\mu\}
\end{eqnarray} 
with non-Abelian field strengths
\beqa
F_R^{\mu\nu} &=& \partial^\mu r^\nu - \partial^\nu r^\mu -
i[r^\mu,r^\nu] \nn
F_L^{\mu\nu} &=& \partial^\mu \ell^\nu - \partial^\nu \ell^\mu -
i [\ell^\mu,\ell^\nu] 
\end{eqnarray}
and with $D_\mu \chi= \partial_\mu \chi - i r_\mu \chi + i\chi l_\mu$.
The covariant derivative
\begin{equation} 
\nabla_\mu X = \partial_\mu X + [\Gamma_\mu,X]
\end{equation} 
is defined in terms of the chiral connection
\begin{equation} 
\Gamma_\mu = \frac{1}{2} \{ u^\dg (\partial_\mu - i r_\mu)u + 
u (\partial_\mu - i \ell_\mu) u^\dg \}~.\label{eq:conn} 
\end{equation} 
With traceless matrix gauge fields $v_\mu$, $a_\mu$, 
the matrices $u_\mu$, $f_\pm^{\mu\nu}$, $h_{\mu\nu}$ and
$\nabla_\lambda f_\pm^{\mu\nu}$ are also traceless. The antisymmetric
combination $\nabla_{[\mu}u_{\nu]}$ is not an independent quantity
because of the identity
\begin{equation} 
f_-^{\mu\nu}=\nabla^\nu u^\mu - \nabla^\mu u^\nu ~. 
\end{equation} 
We use $f_-^{\mu\nu}$ and $\chi_\pm^\mu$ in order to have as few terms
as possible for vanishing external fields.

An equivalent basis is the L(eft)R(ight)-basis with
$U=u^2$ in the conventional choice of coset coordinates. 
We have found it more convenient to use operators
transforming as in (\ref{eq:hXh}) for at least two reasons:
\begin{itemize} 
\item Functional integration produces functionals of such fields in
a natural way, in particular the divergent parts \cite{gl85,bce3,bce4}.
\item One of the main tasks in constructing the Lagrangian of $O(p^6)$
is to find a minimal set closed under partial integration. This is
easier to achieve with operators of type (\ref{eq:hXh}) than in the
LR-basis \cite{fs96}.\footnote{To be more precise, a right-right basis
was used in Ref. \cite{fs96}. In principle, the infinities can also be
calculated directly in this basis \cite{bbc}
but we have used the standard method in \cite{bce3,bce4}.}
\end{itemize} 

The construction of mesonic chiral Lagrangians proceeds by writing
down all Lorentz invariant (products of) traces of products of chiral
operators $X$ with the required chiral dimension. This guarantees
chiral symmetry, but one has to
implement in addition the discrete symmetries $P$ and $C$ and
hermiticity of the Lagrangian. This is straightforward with the help
of the transformation properties given in Table \ref{tab:PCHC}.

\renewcommand{\arraystretch}{1.5}
\begin{table}\begin{center}
\vspace{.5cm}
\begin{tabular}{|c|c|c|c|} 
\hline
\hspace{.5cm} operator \hspace{.5cm} & \hspace{1.5cm} $P$ \hspace{1.5cm}
 &\hspace{1cm} $C$ \hspace{1cm} & \hspace{1cm} h.c. \hspace{1cm} \\ \hline
$u_\mu$ & $-\ve(\mu) u_\mu$ & $u_\mu^T$ & $u_\mu$ \\
$h_{\mu\nu}$ & $-\ve(\mu)\ve(\nu) h_{\mu\nu}$ & $h_{\mu\nu}^T$ & 
$h_{\mu\nu}$ \\
$\chi_\pm$ & $\pm \chi_\pm$ & $\chi_\pm^T$ & $\pm \chi_\pm$ \\
$f_\pm^{\mu\nu}$ & $\pm \ve(\mu)\ve(\nu) f_\pm^{\mu\nu}$ & 
$\mp f_\pm^{\mu\nu T}$ & $f_\pm^{\mu\nu}$ \\
\hline
\end{tabular}\end{center}
\caption{\label{tab:PCHC}$P$, $C$ and hermiticity properties of operators
contained in chiral Lagrangians. Space-time arguments are
suppressed and we do not list the derivatives 
$\chi_{\pm \mu}$ , $\nabla_\lambda f_\pm^{\mu\nu}$  separately.
$\ve(0)=-\ve(\mu\neq 0)=1$.}
\end{table}
 
There is a very large number of terms with chiral dimension six that
fulfill all symmetry constraints. However, many of those terms are
linearly dependent. To obtain a minimal set of independent monomials
of $O(p^6)$ for chiral $SU(n)$, we use the following
relations or procedures:
\begin{enumerate} 
\item[i.] Partial integration in the chiral action of  $O(p^6)$;
\item[ii.] EOM for the lowest-order Lagrangian 
(\ref{eq:L2});
\item[iii.] Bianchi identity;
\item[iv.] Contact terms.
\end{enumerate} 

Before discussing these simplifications in more detail, we present in
Table \ref{tab:L6} the complete list of independent monomials of
$O(p^6)$ for chiral $SU(n)$. There are 112 such terms plus three
contact terms that depend only on external fields. We have ordered the
monomials by introducing subsequently  operators containing
external fields $\chi_+$,  $\chi_-$, $f_+^{\mu\nu}$ and $f_-^{\mu\nu}$,
in this order. For practical purposes, there is one exception to this
rule: the terms with six powers of the vielbein field $u_\mu$ are
listed after the introduction of $\chi_+$ and $\chi_-$, but before
terms involving $f_+^{\mu\nu}$. Such terms will only be relevant for
experimentally rather remote processes involving, e.g., six mesons.

As a check for the completeness and linear independence of the 
112 terms in Table \ref{tab:L6}, we have also employed a different 
basis with higher covariant derivatives that
occur naturally in the calculation \cite{bce3,bce4} of the divergence 
functional of $O(p^6)$. We have explicitly constructed the linear
transformation that transforms the two bases into
one another. Table \ref{tab:L6}
also contains the Lagrangians for $n=2$ and 3 where additional
relations exist, as we will discuss in the next section.

The main tools for reducing the Lagrangian of $O(p^6)$ to its minimal
form are partial integration in the action and the EOM. 
Although straightforward in
principle, it is in practice nontrivial with the huge number 
of possible monomials to find a minimal set closed under partial 
integration. In particular, partial integration together with the EOM
allow to reduce all higher-derivative terms to monomials involving
at most single-derivative operators. It is not very illuminating to
write down all possible relations of this type but we will demonstrate
the procedure with some explicit examples in App.~\ref{app:FS}. In
fact, the optimal use of partial integration is one of the reasons why 
we arrive at a smaller number of independent terms than Ref.~\cite{fs96}.
With partial integration and application of the EOM, there are still 
117 seemingly independent monomials of $O(p^6)$.

The loop expansion can be viewed as an expansion around the
classical solution, i.e., the solution of the EOM. For a
systematic chiral counting,
one expands around the EOM of the lowest-order Lagrangian
(\ref{eq:L2}):
\begin{equation} 
\nabla^\mu u_\mu = \frac{i}{2}\left(\chi_- - \frac{1}{n}\lgl \chi_-
\rgl \right)~.\label{eq:EOM}
\end{equation} 
The generating functional of $O(p^6)$ contains the action for
the Lagrangian of $O(p^6)$ precisely at the classical solution. We may 
therefore replace all occurrences of $\nabla^\mu u_\mu$ or
$h^\mu_{\,\mu}$ by the EOM. As is known to many practitioners in 
quantum field theory, application of the EOM is equivalent to field 
transformations, in our case of the matrix field $u(\vp)$. We include 
a general proof of this equivalence in App.~\ref{app:EOM}.

The chiral coset space has a geometric structure encoded in the
connection (\ref{eq:conn}). The field strength tensor
$\Gamma_{\mu\nu}$ associated with this connection is given by
\begin{eqnarray} 
\left[\nabla_\mu,\nabla_\nu\right]X &=&
\left[\Gamma_{\mu\nu},X\right]\nn
\Gamma_{\mu\nu} &=& \frac{1}{4}[u_\mu,u_\nu]-\frac{i}{2}f_{+\mu\nu}
\end{eqnarray} 
and obeys the Bianchi identity
\begin{equation} 
\nabla_\mu \Gamma_{\nu\rho}+\nabla_\nu \Gamma_{\rho\mu}+
\nabla_\rho \Gamma_{\mu\nu} = 0~.
\end{equation} 
Since the Bianchi identity is of $O(p^3)$, it will lead to
relations among $p^6$ monomials when traced with additional
chiral operators of chiral dimension three. Moreover, 
$\nabla_\rho \Gamma_{\mu\nu}$ is a third-rank Lorentz tensor of even
intrinsic parity. This implies that the Bianchi identity can only
give rise to nontrivial relations when traced with either
$h_{\mu\nu}u_\rho$, $f_{-\mu\nu}u_\rho$ or 
$\nabla_\rho f_{+\mu\nu}$. It turns out that there are only two 
independent relations among the 117 monomials for general $n$ 
when partial integration and
the EOM are applied. We eliminate in this way the
following two monomials that would otherwise appear in the
general list of Table \ref{tab:L6}:
\begin{eqnarray} 
\lgl f_{+\mu\nu} [\chi_-^\mu ,u^\nu]\rgl & , &
\lgl \nabla^\mu f_{+\mu\nu} \nabla_\rho f_+^{\rho\nu} \rgl  ~.
\end{eqnarray} 
This reduces the number of terms to 115 independent ones.

Finally, we turn to the contact terms. Our basis of operators
tends to conceal the fact that some combinations depend only
on external fields and are therefore not directly accessible
experimentally. It is easier to express those contact terms in the
LR-basis. For general $n$, there are three independent contact terms
listed in Table \ref{tab:L6} as entries 113, 114 and 115 
in the $SU(n)$ column. The covariant derivative $D$ contains only 
external gauge fields depending on which object it acts.
For the derivation, we again used partial integration and also
the Bianchi identities for $F^{\mu\nu}_L$ and $F^{\mu\nu}_R$ alone.
Of course,
these three contact terms can be written as linear combinations of
$O(p^6)$ monomials in our basis. We eliminate the
following three monomials in terms of the contact terms and of other
terms contained in the $SU(n)$ list of Table \ref{tab:L6}:
\begin{eqnarray} 
\lgl \chi_{- \mu} \chi_-^\mu \rgl &=& Y_{47} -4 \,Y_{113} \nn 
i\lgl f_{+\mu\nu} [f_+^{\nu\rho},f^\mu_{+\rho}]\rgl &=&
- 3\,Y_{101} - 8\,Y_{114} \nn
\lgl \nabla_\rho f_{+\mu\nu} \nabla^\rho f_+^{\mu\nu} \rgl &=&
3/2\,Y_{71} - 3/2\,Y_{73} - 4\,Y_{75} + 2\,Y_{78} \nn
&&- 1/2\,Y_{90} + 1/2\,Y_{92} - 2\,Y_{100} + 2\,Y_{101} \nn
&&+ 1/2\,Y_{104} -\,Y_{109} - 4\,Y_{111} + 2\,Y_{115} ~.
\end{eqnarray} 
Here and in the following, $Y_i$ stands for the 
$i$-th monomial in the $SU(n)$ column of Table \ref{tab:L6}.

\section{Chiral Lagrangians for $n=2,3$}\label{sec:su23}

The chiral Lagrangian of $O(p^6)$ contains 112 independent terms plus
three contact terms for general $n$. Of course, for phenomenological 
applications  only $n=2$ and 3 are directly relevant. 

For $n=3$, there are additional linear relations among the invariants
of $O(p^6)$ due to the 
Cayley-Hamilton theorem whereby any $n$-dimensional matrix obeys its
own characteristic equation. For $n=3$, this relation implies the 
following identity among three arbitrary three-dimensional matrices $A,B,C$:
\begin{eqnarray} 
\label{eq:CHSU3}
 ABC  +  ACB  +  BAC  + BCA + CAB  +  CBA  & & \\
-  AB  \lgl C \rgl -  AC  \lgl B \rgl 
-  BA  \lgl C \rgl -  BC  \lgl A \rgl 
-  CA  \lgl B \rgl -  CB  \lgl A \rgl & & \nn
-  A  \lgl BC \rgl -  B  \lgl AC \rgl 
-  C  \lgl AB \rgl - \lgl ABC \rgl 
- \lgl ACB \rgl  \nn
+  A  \lgl B \rgl \lgl C \rgl +  B  \lgl A \rgl \lgl C
 \rgl +  C  \lgl A \rgl \lgl B \rgl 
+ \lgl A \rgl \lgl BC 
\rgl    + \lgl B \rgl \lgl AC \rgl
+ \lgl C \rgl \lgl AB \rgl   & & \nn
-  \lgl A \rgl \lgl B \rgl \lgl C \rgl &=& 0 ~.\no
\end{eqnarray} 

Careful analysis leads to 21 independent
relations of the Cayley-Hamilton variety. Thus, we can dispose of
21 of the $SU(n)$ monomials for $n=3$. The explicit relations are
reproduced in App.~\ref{app:CH}. We have chosen to eliminate preferentially
terms with multiple traces to have a transparent large-$N_c$ counting in the
final $O(p^6)$ lagrangian.

We are not aware of a general proof that the Cayley-Hamilton relations
exhaust all possible linear relations among traces of products of
three-dimensional matrices. Moreover, our matrices have in
general special properties such as being hermitian and/or
traceless. We have therefore investigated whether there could be
additional linear relations among the 94 monomials listed in Table 
\ref{tab:L6} for chiral $SU(3)$. In other words, we have looked for
nontrivial solutions of the linear equation
\begin{equation} 
\sum_{i=1}^{94} x_i O_i = 0 \label{eq:lineq}
\end{equation} 
where the $O_i$ denote the 94 monomials relevant for $n=3$.
Of course,  Eq.~(\ref{eq:lineq}) decomposes into several 
independent equations with identical building blocks in the $O_i$.
For example, the monomials $O_{40}$,\ldots,$O_{47}$ can be written as sums
of scalar products of power six of eight real vectors
$u_\mu^a$ ($a=1,\ldots,8$) with  $u_\mu=u_\mu^a\lambda^a$. Without further
restrictions on these vectors, $O_{40}$,\ldots,$O_{47}$ turn out to be
linearly independent.
In general the result of the analysis is that there are indeed no additional
relations for $n=3$ beyond Cayley-Hamilton. In other words, the 
unique solution of (\ref{eq:lineq}) is $x_i=0$ for all $i=1, \dots, 
94$.

One immediate bonus of this analysis is that we obtain all additional
linear relations for chiral $SU(2)$ by restricting
Eq.~(\ref{eq:lineq}) to  two dimensions. The resulting 37
independent relations are listed in App.~\ref{app:CH}. They can all be
interpreted as consequences of the two-dimensional Cayley-Hamilton
theorem which implies the relation
\begin{equation} 
\{ A,B \} = A \lgl B \rgl + B \lgl A \rgl + \lgl AB \rgl - 
\lgl A \rgl\lgl B \rgl 
\end{equation} 
for arbitrary two-dimensional matrices $A,B$.
The most direct way to verify those relations is of course
by making use of the algebra of Pauli matrices.

In addition to the three contact terms for general $n$,
there is a kind of contact term which shows up at different chiral orders,
depending on $n$. This is\footnote{We are indebted to Bachir Moussallam for
pointing out that we had overlooked this contact term for $n=3$
in the original version of the paper.} 
$ \mbox{det}(\chi)+\mbox{det}(\chi^ \dagger)$, which
is a term of order $p^{2n}$. This appears at $O(p^4)$ for $SU(2)$ and at
$O(p^6)$ for $SU(3)$. In the latter case we can express it in our basis
as follows: 
\be 
\mbox{det}(\chi)+\mbox{det}(\chi^\dagger) = {1 \over 12}
Y_{25} - {1 \over 8} Y_{26} + {1 \over 24} Y_{27} + {1 \over 4} Y_{39}
-{1 \over 8} Y_{40} - {1 \over 4} Y_{41}+ {1 \over 8} Y_{42} \; \; .  
\label{eq:detchi}
\ee
We choose to trade $Y_{42}$ for the contact term (\ref{eq:detchi}) in
the $SU(3)$ basis given in Table \ref{tab:L6},
bringing the final count to 90 independent terms $O_i$ plus 4 contact 
terms for chiral $SU(3)$. 

The equivalent of the above term for $n=2$ shows up at $O(p^4)$.
The presence of this contact term can also be explained by
the fact that all representations of $SU(2)$ are self-conjugate:
$\widetilde\chi = \tau_2 \chi^T \tau_2$ transforms like $\chi^\dg$ under
chiral $SU(2)$ transformations, and  $\mbox{det}(\chi)=1/2 \lgl \chi
\widetilde\chi \rgl$. 
Therefore we can also construct a new contact term at $O(p^6)$ for $n=2$ by 
inserting derivatives:
$\lgl D_\mu \chi D^\mu \widetilde\chi \rgl$ + h.c. which is listed at the
end of Table \ref{tab:L6}. We can trade one more invariant
for this contact term by using the relation:
\begin{eqnarray} 
\lgl \chi_{- \mu} \rgl \lgl \chi_-^\mu \rgl  &=&
 \, 2 Y_{47} -\,Y_{48} - \, 4 Y_{113} + 
2 (\lgl D_\mu \chi D^\mu \widetilde\chi \rgl + {\rm ~h.c.}) ~,
\label{eq:ctt}
\end{eqnarray}  
bringing the final count to 53 independent terms $P_i$ plus 4 contact 
terms for chiral $SU(2)$. 

The mesonic chiral Lagrangians of $O(p^6)$ then take the following
final forms:
\begin{eqnarray} 
\cL_6^{SU(3)} &=& \sum_{i=1}^{90} C_i O_i + 4\,\, {\rm contact \,\, 
terms} \nn
\cL_6^{SU(2)} &=& \sum_{i=1}^{53} c_i P_i + 4\,\, {\rm contact \,\,
terms} ~,
\end{eqnarray}
with 90 (53) low-energy constants $C_i$ ($c_i$) for
$n=3~(2)$. These low-energy constants parametrize the most general
local solutions of the chiral Ward identities. Most of them have a
divergent part to cancel the divergences \cite{bce3,bce4} of the one- 
and two-loop functionals of $O(p^6)$. The scale-dependent remainders
$C_i^r(\mu)$ [$c_i^r(\mu)]$ are at least in principle all measurable
quantities.

\section{A guided tour through ${\cal L}_6$}\label{sec:tour}

The number of new constants appearing at order $p^6$ is very large. 
One cannot hope to determine all of them from experiments,
as could be done with few additional assumptions for $\cL_4$.
On the other hand, a closer look at the various 
terms appearing in $\cL_6$ shows that a large fraction of them is not
phenomenologically relevant. We allude here to terms like $Y_{49},
\ldots, Y_{63}$ that contribute to processes involving at least six 
mesons, or like $Y_{101}$ that involves at least one vector and two 
axial currents. To help the reader in identifying such terms, the last 
column in Table \ref{tab:L6} shows the simplest quantity or process to 
which each term contributes.

Actually, the situation is not as hopeless as it may
look at first sight. The number of terms relevant for
phenomenological purposes is still manageable.  To better illustrate
this claim, we concentrate here on the Lagrangian for $n=2$
and in the chiral limit, for $s=p=0$ and $v_\mu, a_\mu \ne 0$. This is 
practically all one needs for phenomenological applications, since the
external sources $s$ and $p$ are not realized in nature, and because 
moving away from the chiral limit in the $u$, $d$ sector produces only 
a small effect. In other words, when
we include all terms proportional to $M^2=B(m_u+m_d)$ we only add 
small corrections to a momentum structure that is already
present at $O(p^4)$, renormalizing the corresponding low--energy
constants.

Restricting ourselves to this simplified situation and to processes
with at most four pions or currents (with not more than one axial
current), the number of
phenomenologically relevant terms goes down to sixteen:
\begin{eqnarray}
P_1,\; P_2, \; P_3 && \mbox{contributing to} \; \; \pi \pi \to
\pi \pi, \nn 
P_{29},\ldots , P_{33}, P_{50} && \mbox{contributing to} \; \; 
\gamma \gamma \rightarrow \pi \pi , \nn  
P_{36}, \; P_{37}, \; P_{38} && \mbox{contributing to} \; \; \tau
\rightarrow 3 \pi \nu_\tau , \nn
P_{44}, \; P_{50} && \mbox{contributing to} \; \; \pi \to l \nu \gamma, 
\nn
P_{51}, \; P_{53} && \mbox{contributing to} \; \; F_V^\pi(t),\nn
P_{52} && \mbox{contributing to} \; \; \pi \to l \nu \gamma^* . 
\nonumber
\end{eqnarray}
This is indeed a more manageable number of coupling constants. Still,
the real difficulty here is to relate different observables
in a useful and practicable manner. 

The Lagrangian we have constructed shows that at $O(p^6)$
chiral symmetry becomes much less restrictive than in lower orders. In
other words, the corresponding Ward identities involve
a large number of different observables. Either one finds a way to
calculate, or at least estimate, the various constants appearing at
this order, or the use of this Lagrangian for relating different 
observables via chiral symmetry will be rather limited.

\section{Conclusion}\label{sec:conc}

We have constructed the most general even-intrinsic-parity
Lagrangian of order
$p^6$ in the mesonic sector for the strong interaction in the presence
of external vector, axial-vector, scalar and pseudoscalar fields. We used
partial integration in the action, the equations of motion, or equivalently
field redefinitions, and Bianchi identities in order to reduce the number
of terms to a minimal set.
We presented a general proof for the equivalence between EOM and
field redefinitions for eliminating spurious terms in chiral 
Lagrangians.

The Lagrangian contains 112 in principle measurable terms and 
3 contact terms for
the general case of $n$ light flavours. For $n=3$ (2), the 
Cayley-Hamilton relations reduce the respective Lagrangians to
90 (53) measurable terms and 4 (4) contact terms. The
differences between our result and earlier ones were discussed.

The chiral Lagrangian of $O(p^6)$ is a necessary requisite for
the renormalization program at the two-loop level 
\cite{bce3,bce4}. The low-energy constants of this Lagrangian 
parametrize the most general local solution of $O(p^6)$ of the 
chiral Ward identities. Although it will not be possible to
determine all renormalized low-energy constants by comparison with
experiment, we now have a basis at our disposal for investigating
those coupling constants with additional theoretical input beyond 
pure symmetry considerations (resonance saturation, large-$N_c$,
lattice simulations, chiral models, \dots ).

\vspace*{0.5cm}

\acknowledgments
We thank J.~Gasser for useful comments and constant encouragement.
We are also grateful to
H.W. Fearing and H. Neufeld for comments on the manuscript.
Finally, we thank B. Moussallam for pointing out the existence of a
fourth contact term for $n=3$. 

\appendix
\setcounter{equation}{0}
\newcounter{zahler}
\addtocounter{zahler}{1}
\renewcommand{\thesection}{\Alph{zahler}}
\renewcommand{\theequation}{\Alph{zahler}.\arabic{equation}}
\section{Field redefinitions and equations of motion}
\label{app:EOM}

In this appendix we give a general proof for the equivalence 
of two procedures for removing terms in effective chiral Lagrangians:
\begin{enumerate} 
\item Using the lowest-order EOM; 
\item Performing field redefinitions (FR).
\end{enumerate} 
The following proof completes the discussion of Ref.~\cite{fs95} 
where it was explicitly shown that EOM terms can be removed by FR. 
 
In full generality, an element of the chiral coset space is of the
form $(l_L(\phi),l_R(\phi))$. Under a chiral transformation 
$g=(g_L,g_R) \in SU(n)_L \times SU(n)_R$,
\begin{equation} 
l_A(\phi) \to g_A l_A(\phi) h(g,\phi)^{-1} \qquad A=L,R
~.\label{eq:lLR} 
\end{equation}
Parity as an automorphism of the chiral group interchanges $l_L$ and
$l_R$. The matrix field $U(\phi)$ used in the LR-basis is defined as
\begin{equation} 
U(\phi)=l_R(\phi) l_L(\phi)^\dg ~.
\end{equation} 
The usual choice of coset coordinates \cite{ccwz} corresponds to
\begin{eqnarray} 
l_R(\phi)&=& l_L(\phi)^\dg =: u(\phi) \nn
U(\phi)&=&u(\phi)^2 ~.
\end{eqnarray} 

Let us now assume that the general chiral Lagrangian has been
constructed in this standard basis including all EOM terms
(external fields are denoted collectively as $j$):
\begin{eqnarray} 
\label{eq:Leff}
\cL(\phi,j) &=& \cL_2(\phi,j) + \sum_{n\ge2}\left(\cL_{2n}(\phi,j) +
\lgl X_2(\phi,j) \cE_{2n-2}(\phi,j)\rgl \right)~. 
\end{eqnarray}  
The last terms are the EOM terms of chiral order $2n$ with
\begin{equation} 
X_2(\phi,j) = \nabla^\mu u_\mu - \frac{i}{2}\left(\chi_- - 
\frac{1}{n}\lgl \chi_- \rgl \right)~.\label{eq:X2}
\end{equation}

We now perform a FR which amounts to a reparametrization of coset
space. The most general such transformation is of the form
\begin{equation} 
\hat l_R(\phi,j) = u(\phi) e^{\frac{i}{2}\xi(\phi,j)}
e^{i\sigma(\phi,j)}~.\label{eq:gFR}
\end{equation} 
Because $\hat l_R$ and $u$ are elements of $SU(n)$, the matrix fields 
$\xi$ and $\sigma$ are hermitian and traceless. We have split the
transformation matrix into two parts distinguished by their intrinsic
parity: $\sigma$ is even while $\xi$ is odd. Parity then fixes 
$\hat l_L$ to be
\begin{equation} 
\hat l_L(\phi,j) = u(\phi)^\dg e^{-\frac{i}{2}\xi(\phi,j)}
e^{i\sigma(\phi,j)}~.
\end{equation} 

For showing the equivalence between EOM and FR, it is important to 
realize that $\sigma(\phi,j)$ has no effect on the chiral Lagrangian. 
This is most easily seen by looking at the matrix $\hat U$ after the
FR: 
\begin{eqnarray} 
\hat U &=& \hat l_R \hat l_L^\dg = u e^{\frac{i}{2}\xi}
e^{i\sigma}e^{-i\sigma}e^{\frac{i}{2}\xi} u 
= u e^{i\xi} u ~.
\end{eqnarray} 
Therefore, the most general FR is given by a hermitian, traceless
matrix field $\xi(\phi,j)$ of odd intrinsic parity. Moreover, from the
definition (\ref{eq:gFR}) and the transformation
property (\ref{eq:lLR}), the field $\xi(\phi,j)$ transforms as in
(\ref{eq:hXh}). Since $X_2$ in
(\ref{eq:X2}) shares these properties, this also holds for the
matrices $\cE_{2n}(\phi,j)$ in (\ref{eq:Leff}). In particular,
$\cE_{2n}(\phi,j)$ can be taken to be traceless in all generality
since the trace of $\cE_{2n}(\phi,j)$ does not contribute in
(\ref{eq:Leff}). Therefore, each
term in $\cE_{2n}(\phi,j)$ defines a possible FR via $\xi(\phi,j)$
and vice versa.

The remainder of the proof is as in Ref.~\cite{fs95}. In a first step,
we choose $\xi=\xi_2$ of $O(p^2)$ as
\begin{equation} 
\xi_2(\phi,j) = - \frac{2}{F^2} \cE_2(\phi,j) ~.
\end{equation} 
After this FR, the lowest-order Lagrangian (\ref{eq:L2}) turns into
(including partial integration in the action)
\begin{equation} 
\cL_2 \to \cL_2 + \frac{F^2}{2}\lgl \xi_2 X_2 \rgl + O(p^6) = \cL_2 -
\lgl X_2 \cE_2 \rgl + O(p^6),
\end{equation} 
canceling the EOM terms of $O(p^4)$ in (\ref{eq:Leff}). The chiral
Lagrangian of $O(p^6)$ and higher (including the EOM terms) will of
course be modified. Since we are considering the most general chiral 
Lagrangian in (\ref{eq:Leff}), only the coefficients of the higher-order
terms may have changed, but not their structure.

We can now repeat the procedure by introducing a further FR
\begin{equation} 
\xi_4(\phi,j) = - \frac{2}{F^2} \cE_4(\phi,j) 
\end{equation} 
that removes the EOM terms of $O(p^6)$ in (\ref{eq:Leff}) without
modifying $\cL_2$ and $\cL_4$, but affecting of course the Lagrangian
of $O(p^8)$ and higher. Continuing this procedure, we can remove all
EOM terms order by order in $p^2$ by successive FR of the form
\begin{equation} 
\hat l_R(\phi,j) = u(\phi) e^{\frac{i}{2}\xi_2(\phi,j)}
\dots e^{\frac{i}{2}\xi_{2n}(\phi,j)} \dots ~.
\end{equation} 

Since the most general expression for $\cE_{2n}(\phi,j)$ defines the
most general possible FR in terms of $\xi_{2n}(\phi,j)$ and vice
versa, the equivalence between using the EOM and performing FR is
established.

\setcounter{equation}{0}
\addtocounter{zahler}{1}

\section{Cayley-Hamilton relations}
\label{app:CH}

The Cayley-Hamilton theorem for $n=3$ can be used to eliminate the
following 21 terms from the $SU(n)$ list. As emphasized in the
introduction, we eliminate preferentially multiple trace terms
(same for the $n=2$ relations below) to make the large-$N_c$ structure
manifest. Here as in the main text, $Y_i$ stands for the 
$i$-th monomial in the $SU(n)$ column of Table \ref{tab:L6}.
The left-hand side mentions in square brackets also the numbering of the
monomial removed. We use the notation $u\ccdot u=u_\rho u^\rho$.
\begin{eqnarray} 
\label{eq:relsu3}
\lgl h_{\mu \nu} u_\rho \rgl \lgl h^{\mu\nu} u^\rho \rgl
[Y_4]&=& 2\,Y_{1} - 1/2\,Y_{2} +\,Y_{3} \nn
\lgl h_{\mu \nu} u_\rho \rgl \lgl h^{\mu\rho} u^\nu \rgl [Y_6]&=&
2/3\,Y_{1} - 2/3\,Y_{3} +\,Y_{5} +\,Y_{28} - 1/2\,Y_{29} \nn 
&&- 1/3\,Y_{30} -\,Y_{31} + 2/3\,Y_{33} - 2/3\,Y_{37} - 4/3\,Y_{49} \nn
&&+ 1/3\,Y_{50} - 2/3\,Y_{52} + 8/3\,Y_{54} - 1/3\,Y_{57} + 2/3\,Y_{58} \nn
&&- 4/3\,Y_{60}  - 2/3\,Y_{64} + 2/3\,Y_{65} - 4/3\,Y_{66} - 8/3\,Y_{67} \nn
&&+ 2/3\,Y_{68} + 4/3\,Y_{86} - 2/3\,Y_{87} - 4/3\,Y_{88} + 2/3\,Y_{89} \nn
&&- 2/3\,Y_{90} + 2/3\,Y_{92} + 2\,Y_{94} - 2/3\,Y_{95} - 1/3\,Y_{96} \nn
&&- 2/3\,Y_{97} + 1/3\,Y_{99} + 2/3\,Y_{105} + 1/3\,Y_{106}\nn
\lgl u\ccdot u \rgl^2 \lgl \chi_+ \rgl [Y_{10}]&=&
 - 4\,Y_{7} + 2\,Y_{8} + 3\,Y_{9} - 2\,Y_{11} + 2\,Y_{12} \nn
\lgl \chi_+ u_\mu u_\nu \rgl \lgl u^\mu u^\nu \rgl [Y_{15}]&=&
Y_{7} - 1/2\,Y_{9} +\,Y_{11} -\,Y_{12} +\,Y_{13} \nn
 \lgl \chi_+ \rgl \lgl u_\mu u_\nu \rgl^2 [Y_{16}]&=&
2\,Y_{7} +\,Y_{8} - 3/2\,Y_{9} +\,Y_{11} -\,Y_{12} +\,Y_{14} \nn
\lgl u\ccdot u \rgl \lgl \chi_+ \rgl^2 [Y_{22}]&=&
- 4\,Y_{19} + 4\,Y_{20} +\,Y_{21} - 2\,Y_{23} + 2\,Y_{24} \nn
i \lgl h_{\mu \nu} u^\mu \rgl \lgl \chi_- u^\nu \rgl [Y_{32}]&=&
Y_{28} - 1/2\,Y_{29} -\,Y_{30} +\,Y_{31} \nn
\lgl u\ccdot u \rgl \lgl \chi_- \rgl^2 [Y_{36}]&=&
 - 4\,Y_{33} + 4\,Y_{34} +\,Y_{35} - 2\,Y_{37} + 2\,Y_{38} \nn
\lgl u\ccdot u \rgl^3 [Y_{51}]&=&
 - 4\,Y_{49} + 5\,Y_{50} - 2\,Y_{52} + 2\,Y_{53} \nn
 \lgl u\ccdot u u_\mu u_\nu \rgl \lgl u^\mu u^\nu \rgl [Y_{55}]&=&
Y_{49} - 1/2\,Y_{50} +\,Y_{52} -\,Y_{53} +\,Y_{54} \nn
\lgl u\ccdot u \rgl \lgl u_\mu u_\nu \rgl^2 [Y_{56}]&=&
 2\,Y_{49} - 1/2\,Y_{50} +\,Y_{52} -\,Y_{53} +\,Y_{57} \nn
\lgl u_\mu u_\nu u_\rho \rgl^2 [Y_{59}]&=&
-\,Y_{49} + 3/4\,Y_{50} - 3/2\,Y_{52} + 3/2\,Y_{53} +\,Y_{58} \nn
\lgl u_\mu u_\nu u_\rho \rgl \lgl u^\mu u^\rho u^\nu \rgl [Y_{61}]&=&
 -\,Y_{49} + 1/4\,Y_{50} + 1/2\,Y_{52} + 3/2\,Y_{53} \nn
&&- 2\,Y_{54} + 1/2\,Y_{57} +\,Y_{60} \nn
\lgl u_\mu u_\nu \rgl \lgl u_\rho u^\mu u^\rho u^\nu \rgl [Y_{62}]&=&
 -\,Y_{53} + 2\,Y_{54} - 1/2\,Y_{57} +\,Y_{60} \nn
\lgl u_\mu u_\nu \rgl \lgl u^\mu u_\rho \rgl \lgl u^\nu u^\rho \rgl 
[Y_{63}]&=&
\,Y_{49} - 3/4\,Y_{50} + 3/2\,Y_{52} - 5/2\,Y_{53} + 4\,Y_{54} -\,Y_{57}
 +\,Y_{60} \nn
i \lgl f_{+ \mu \nu} u_\rho \rgl \lgl  u^\mu u^\nu u^\rho \rgl [Y_{69}]&=&
\,Y_{64} - 1/2\,Y_{65} +\,Y_{66} \nn
i \lgl f_{+ \mu \nu} [u^\mu ,u_\rho ] \rgl \lgl u^\nu u^\rho \rgl [Y_{70}]&=&
\,Y_{64} -\,Y_{65} + 2\,Y_{67} +\,Y_{68} \nn
\lgl f_{+ \mu \nu} u_\rho \rgl^2 [Y_{74}]&=&2\,Y_{71} - 1/2\,Y_{72} +\,Y_{73}
 \nn
\lgl f_{+ \mu \nu} u_\rho \rgl \lgl f_+^{\mu\rho} u^\nu \rgl [Y_{79}]&=&
 2\,Y_{75} + 2\,Y_{76} -\,Y_{77} +\,Y_{78} -\,Y_{80} \nn
\lgl f_{- \mu \nu} u_\rho \rgl^2 [Y_{93}]&=& 2\,Y_{90} - 1/2\,Y_{91} +\,Y_{92}
 \nn
\lgl f_{- \mu \nu} u_\rho \rgl \lgl f_-^{\mu\rho} u^\nu \rgl [Y_{98}]&=&
 2\,Y_{94} + 2\,Y_{95} -\,Y_{96} +\,Y_{97} -\,Y_{99} ~.
\end{eqnarray} 

All these relations can be derived from the identity
(\ref{eq:CHSU3}) by multiplying with appropriate chiral matrices and
taking traces. Only the second relation of (\ref{eq:relsu3}) 
needs some more explanation.
The original relation takes the simpler form
\begin{equation} 
Y_{86}-Y_{87}-Y_{88}+Y_{89}+ \lgl f_{-\mu\nu}(h^{\nu\rho}u^\mu u_\rho
+ u_\rho u^\mu h^{\nu\rho})\rgl + \lgl f_{-\mu\nu} u_\rho \rgl
\lgl h^{\nu\rho}u^\mu \rgl = 0 ~.
\end{equation} 
However, we have already replaced the last two terms in this relation
in the $SU(n)$ basis by partial integration and EOM leading to the
more complicated form of the second relation in (\ref{eq:relsu3}).

In addition, the extra contact term has been used to remove $Y_{42}$
via (\ref{eq:detchi}).

As explained in Sect.~2, there are 37 additional linear relations for
$n=2$ that can be written in the form
\begin{eqnarray} 
\label{eq:relsu2}
\lgl u\ccdot u \rgl \lgl h_{\mu \nu} h^{\mu\nu} \rgl 
[Y_{2}]&=& 2\,Y_{1} \nn
\lgl (u\ccdot u)^2 \rgl \lgl \chi_+ \rgl 
[Y_{8}]&=& 2\,Y_{7} \nn
\lgl u\ccdot u \rgl \lgl u\ccdot u \chi_+ \rgl 
[Y_{9}]&=& 2\,Y_{7} \nn
\lgl u\ccdot u u_\mu \chi_+ u^\mu \rgl 
[Y_{11}]&=&\,Y_{7} \nn
\lgl u\ccdot u u_\mu \rgl \lgl \chi_+ u^\mu \rgl 
[Y_{12}]&=& 0 \nn
\lgl \chi_+ \rgl \lgl  u_\mu u_\nu u^\mu u^\nu \rgl 
[Y_{14}]&=& 2\,Y_{13} \nn
\lgl \chi_+ \rgl \lgl h_{\mu \nu} h^{\mu\nu} \rgl 
[Y_{18}]&=& 2\,Y_{17} \nn
\lgl u\ccdot u \rgl \lgl \chi_+^2 \rgl 
[Y_{21}]&=& 2\,Y_{19} \nn
\lgl \chi_+ u_\mu \rgl^2 
[Y_{24}]&=&\,Y_{19} -\,Y_{20} +\,Y_{23} \nn
\lgl \chi_+ \rgl^3 
[Y_{27}]&=& - 2\,Y_{25} + 3\,Y_{26} \nn
i \lgl \chi_-  h_{\mu \nu} \rgl \lgl u^\mu u^\nu \rgl 
[Y_{29}]&=&\,Y_{28} \nn
i \lgl h_{\mu \nu} u^\mu u^\nu \rgl \lgl \chi_- \rgl 
[Y_{30}]&=& 0 \nn
\lgl u\ccdot u \rgl \lgl \chi_-^2 \rgl 
[Y_{35}]&=& 2\,Y_{33} \nn
\lgl u_\mu \chi_- \rgl^2 
[Y_{38}]&=&\,Y_{33} -\,Y_{34} +\,Y_{37} \nn
\lgl \chi_+ \rgl \lgl \chi_- \rgl^2 
[Y_{42}]&=& - 2 \,Y_{39}+\,Y_{40} + 2 \,Y_{41} \nn
i \lgl \chi_{+ \mu} \rgl \lgl \chi_- u^\mu \rgl 
[Y_{45}]&=&\,Y_{43} -\,Y_{44} \nn
\lgl (u\ccdot u)^2 \rgl \lgl u\ccdot u \rgl 
[Y_{50}]&=& 2\,Y_{49} \nn
\lgl u\ccdot u u_\mu u\ccdot u u^\mu \rgl 
[Y_{52}]&=&\,Y_{49} \nn
\lgl u\ccdot u u_\mu \rgl^2 
[Y_{53}]&=& 0 \nn
\lgl u\ccdot u \rgl \lgl u_\mu u_\nu u^\mu u^\nu \rgl 
[Y_{57}]&=& 2\,Y_{54} \nn
\lgl u_\mu u_\nu u_\rho u^\mu u^\rho u^\nu \rgl 
[Y_{60}]&=&\,Y_{49} +\,Y_{54} -\,Y_{58} \nn
i \lgl f_{+ \mu \nu} \{u\ccdot u, u^\mu u^\nu \} \rgl 
[Y_{64}]&=& 2\,Y_{67} \nn
i \lgl u\ccdot u \rgl \lgl f_{+ \mu \nu} u^\mu u^\nu \rgl 
[Y_{65}]&=& 2\,Y_{67} \nn
i \lgl f_{+ \mu \nu} \{ u_\rho , u^\mu u^\rho u^\nu \} \rgl 
[Y_{68}]&=&  -\,Y_{66} -\,Y_{67} \nn
\lgl u\ccdot u \rgl \lgl f_{+ \mu \nu} f_+^{\mu\nu} \rgl
[Y_{72}]&=& 2\,Y_{71} \nn
\lgl f_{+ \mu \nu} f_+^{\mu\rho} \rgl \lgl u^\nu u_\rho \rgl
[Y_{77}]&=&\,Y_{75} + Y_{76} \nn 
\lgl f_{+ \mu \nu} u^\nu \rgl \lgl f_+^{\mu\rho} u_\rho \rgl 
[Y_{80}]&=& Y_{76} + 1/2\,Y_{78} \nn
\lgl \chi_+ \rgl \lgl f_{+ \mu \nu} f_+^{\mu\nu} \rgl 
[Y_{82}]&=& 2\,Y_{81} \nn
i \lgl f_{+ \mu \nu} \{\chi_+, u^\mu u^\nu \} \rgl 
[Y_{83}]&=& 2\,Y_{85} \nn
i \lgl \chi_+ \rgl \lgl f_{+ \mu \nu} u^\mu u^\nu \rgl 
[Y_{84}]&=& 2\,Y_{85} \nn
\lgl f_{- \mu \nu} u^\mu \rgl \lgl h^{\nu\rho} u_\rho \rgl 
[Y_{88}]&=& 1/2\,Y_{86} + 1/2\,Y_{89} \nn
\lgl u\ccdot u \rgl \lgl f_{- \mu \nu} f_-^{\mu\nu} \rgl 
[Y_{91}]&=& 2\,Y_{90} \nn
\lgl f_{- \mu \nu} f_-^{\mu\rho} \rgl \lgl u^\nu u_\rho \rgl 
[Y_{96}]&=&  Y_{94} +\,Y_{95} \nn
\lgl f_{- \mu \nu} u^\nu \rgl \lgl f_-^{\mu\rho} u_\rho \rgl 
[Y_{99}]&=& \,Y_{95} + 1/2\,Y_{97} \nn
\lgl \chi_+ \rgl \lgl f_{- \mu \nu} f_-^{\mu\nu} \rgl 
[Y_{103}]&=& 2\,Y_{102} \nn
i \lgl f_{- \mu \nu} u^\nu \rgl \lgl u^\mu \chi_- \rgl 
[Y_{106}]&=&\,Y_{105}/2 \nn
\lgl \chi_+^\mu \rgl \lgl f_{- \mu \nu} u^\nu \rgl 
[Y_{108}]&=&\,Y_{107} ~.
\end{eqnarray}

In addition, the extra contact term has been used to remove $Y_{46}$
via (\ref{eq:ctt}).

\setcounter{equation}{0}
\addtocounter{zahler}{1}

\section{Comparison with the Lagrangian of \\Fearing and Scherer}
\label{app:FS}

Fearing and Scherer (FS) \cite{fs96} have constructed chiral Lagrangians 
of $O(p^6)$ in the LR-basis, but otherwise they have followed a
strategy very similar to ours. We have explicitly checked that all
their terms can be expressed in our basis of independent chiral
invariants of $O(p^6)$. However, since they
have\footnote{In \cite{fs96} no attempt was made
to create a minimal basis for the $n$-flavour case. They quote 18 trace
relations and 111 terms for 3 flavours, hence we use 129 as their number
of terms for $n$ flavours.} 129 instead of our 112+3 monomials for $SU(n)$ and
111 vs. 90+4 for $SU(3)$, their basis can not be minimal.

Two obvious reasons for these differences are the Bianchi identity and
the contact terms, both of which FS did not take into
consideration. Nevertheless, this does not fully account for the different
number of terms, which can then only be due to partial integration
and/or EOM for general $n$ and to the Cayley-Hamilton theorem for $n=3$.

We now exhibit some examples of linear dependences in the FS 
Lagrangian. Since we do not intend to provide a full translation of
their 129 (111) terms into our basis, we restrict the comparison to
the case where all external fields are set to zero. In our basis, this
leaves 21 terms denoted $Y_1, \dots, Y_6$ and $Y_{49}, \dots, Y_{63}$
for chiral $SU(n)$.
In comparison, FS also have the equivalent of  $Y_{49}, \dots,
Y_{63}$, but 8 additional terms instead of our $Y_1, \dots, Y_6$.
Four of their additional terms are directly related to $Y_1, Y_2, Y_4,
Y_6$ (and to $Y_{49}, \dots, Y_{63}$). The other four (in our notation 
and with all external fields set
to zero) can be expressed in our basis 
via partial integration and application of the EOM:
\begin{eqnarray} 
\lgl h_{\mu \nu} h^{\mu\rho} u^\nu u_\rho \rgl &=&
       - 1/2\,Y_{3}  +\,Y_{54} -\,Y_{60} \nn 
\lgl h_{\mu \nu} h^{\mu\rho} u_\rho u^\nu \rgl &=&
       -\,Y_{1} - 1/2\,Y_{5}  + \,Y_{49} +\,Y_{52} - 2\,Y_{54} \nn 
\lgl h_{\mu \nu} h^{\mu\rho} \rgl \lgl u^\nu u_\rho \rgl &=&
       -\,Y_{4} -\,Y_{6}  + 2\,Y_{55} - 2\,Y_{62} \nn
\lgl h_{\mu \nu} u^\nu \rgl \lgl h^{\mu \rho} u_\rho \rgl &=&
       - 1/2\,Y_{2}  +\,Y_{50} -\,Y_{57} ~.
\end{eqnarray} 
Thus, for general $n$ there are two linear relations among the 
23 terms used by FS.

Turning now to $SU(3)$, FS have 12 terms with six powers of $u_\mu$
while we have only eight. We have six of the $Y_i$ in common
(i=49,50,52,53,54,58). The remaining six invariants of FS can be
expressed with the help of Cayley-Hamilton in terms of our
eight invariants that also include $Y_{57}$ and $Y_{60}$. The
following equalities already appear in (\ref{eq:relsu3}).
\begin{eqnarray} 
\lgl u\ccdot u \rgl^3 &=&
 - 4\,Y_{49} + 5\,Y_{50} - 2\,Y_{52} + 2\,Y_{53} \\
 \lgl u\ccdot u u_\mu u_\nu \rgl \lgl u^\mu u^\nu \rgl &=&
Y_{49} - 1/2\,Y_{50} +\,Y_{52} -\,Y_{53} +\,Y_{54} \nn
\lgl u\ccdot u \rgl \lgl u_\mu u_\nu \rgl^2 &=&
 2\,Y_{49} - 1/2\,Y_{50} +\,Y_{52} -\,Y_{53} +\,Y_{57} \nn
\lgl u_\mu u_\nu u_\rho \rgl^2 &=&
-\,Y_{49} + 3/4\,Y_{50} - 3/2\,Y_{52} + 3/2\,Y_{53} +\,Y_{58} \nn
\lgl u_\mu u_\nu u_\rho \rgl \lgl u^\mu u^\rho u^\nu \rgl &=&
 -\,Y_{49} + 1/4\,Y_{50} + 1/2\,Y_{52} + 3/2\,Y_{53} \nn
&&- 2\,Y_{54} + 1/2\,Y_{57} +\,Y_{60} \nn
\lgl u_\mu u_\nu \rgl \lgl u^\mu u_\rho \rgl \lgl u^\nu u^\rho \rgl 
&=&\,Y_{49} - 3/4\,Y_{50} + 3/2\,Y_{52} - 5/2\,Y_{53} + 4\,Y_{54} 
-\,Y_{57} +\,Y_{60} ~. \no
\end{eqnarray} 
Consequently, there are four linear relations of the Cayley-Hamilton
type among the 12 terms of FS.

\newpage

\renewcommand{\arraystretch}{1.1}
\setlength{\LTcapwidth}{\textwidth}

\begin{longtable}[c]{|lc|c|c|c|c|}

\hline
\hspace{2cm} monomial ($Y_i$) & \hspace{0cm} & SU(n) & SU(3) & SU(2)&
contributes to \\
\hline
\hline
\endhead
\hline
\caption[]{\rule{0cm}{2em}}
\endfoot
\hline
\caption[Independent monomials of  $O(p^6)$ for $SU(n)$,
 $SU(3)$ and $SU(2)$. The flavour trace is denoted by 
$\lgl \dots \rgl$. The first column lists the structure and the
following three the numbering of independent terms for a number of flavours
$n$, 3 and 2. The last column indicates the simplest quantity or
process to which the term contributes.]{\label{tab:L6} 
\rule{0cm}{2em}Independent monomials of  $O(p^6)$ for $SU(n)$,
 $SU(3)$ and $SU(2)$. The flavour trace is denoted by 
$\lgl \dots \rgl$. The first column lists the structure and the
following three the numbering of independent terms for a number of flavours
$n$, 3 and 2. The last column indicates the simplest quantity or
process to which the term contributes.} 
\endlastfoot
$\lgl u\ccdot u h_{\mu \nu} h^{\mu\nu} \rgl$& & 1 & 1&1 & $\pi \pi
\rightarrow \pi \pi$ \\ 
 $\lgl u\ccdot u \rgl \lgl h_{\mu \nu} h^{\mu\nu} \rgl$& & 2 & 2& & $\pi
 \pi \rightarrow \pi \pi$ \\ 
$\lgl h_{\mu \nu} u_\rho h^{\mu\nu} u^\rho \rgl$& & 3 & 3 &2 & $\pi \pi
\rightarrow \pi \pi$ \\ 
 $\lgl h_{\mu \nu} u_\rho \rgl \lgl h^{\mu\nu} u^\rho \rgl$& & 4 & & &
 $\pi \pi \rightarrow \pi \pi$ \\
$\lgl h_{\mu \nu} \left(u_\rho h^{\mu\rho} u^\nu + u^\nu h^{\mu
    \rho} u_\rho \right) \rgl$& & 5 & 4 &3& $\pi \pi \rightarrow \pi \pi$ \\
 $\lgl h_{\mu \nu} u_\rho \rgl \lgl h^{\mu\rho} u^\nu \rgl$& & 6 & & &
 $\pi \pi \rightarrow \pi \pi$ \\
 & &  & & & \\
$\lgl (u\ccdot u)^2 \chi_+ \rgl$& & 7 & 5 &4 & $\pi \pi \rightarrow \pi
\pi$ \\ 
 $\lgl (u\ccdot u)^2 \rgl \lgl \chi_+ \rgl$& & 8 & 6 & & $\pi \pi
 \rightarrow \pi \pi$\\ 
 $\lgl u\ccdot u \rgl \lgl u\ccdot u \chi_+ \rgl$& & 9 & 7 & & $\pi \pi
 \rightarrow \pi \pi$\\ 
 $\lgl u\ccdot u \rgl^2 \lgl \chi_+ \rgl$& & 10 & &  & $\pi \pi \rightarrow
 \pi \pi$ \\ 
 $\lgl u\ccdot u u_\mu \chi_+ u^\mu \rgl$& & 11 & 8 & & $\pi \pi
 \rightarrow \pi \pi$\\ 
 $\lgl u\ccdot u u_\mu \rgl \lgl \chi_+ u^\mu \rgl$& & 12 & 9 & & $\pi \pi
 \rightarrow \pi \pi$\\ 
$\lgl \chi_+ u_\mu u_\nu u^\mu u^\nu \rgl$& & 13 & 10&5 & $\pi \pi
\rightarrow \pi \pi$\\ 
 $\lgl \chi_+ \rgl \lgl  u_\mu u_\nu u^\mu u^\nu \rgl$ & &14 & 11 & &
 $\pi \pi \rightarrow \pi \pi$ \\
 $\lgl \chi_+ u_\mu u_\nu \rgl \lgl u^\mu u^\nu \rgl$& & 15 & & & $\pi \pi
 \rightarrow \pi \pi$ \\ 
 $\lgl \chi_+ \rgl \lgl u_\mu u_\nu \rgl^2$& & 16 & & & $\pi \pi
 \rightarrow \pi \pi$ \\ 
$\lgl \chi_+ h_{\mu \nu} h^{\mu\nu} \rgl$& & 17 & 12 & 6 & $\lgl \pi \pi
\rgl$\\ 
 $\lgl \chi_+ \rgl \lgl h_{\mu \nu} h^{\mu\nu} \rgl$& & 18 & 13 & & $\lgl
 \pi \pi \rgl$\\  
$\lgl u\ccdot u \chi_+^2 \rgl$& & 19 & 14 & 7 & $\lgl \pi \pi \rgl$\\
$\lgl u\ccdot u \chi_+ \rgl \lgl \chi_+ \rgl$& & 20 & 15 & 8 & $\lgl \pi
\pi \rgl$\\ 
 $\lgl u\ccdot u \rgl \lgl \chi_+^2 \rgl$& & 21 & 16 & & $\lgl \pi \pi \rgl$ \\
 $\lgl u\ccdot u \rgl \lgl \chi_+ \rgl^2$& & 22 & & & $\lgl \pi \pi \rgl$ \\
$\lgl \chi_+ u_\mu \chi_+ u^\mu \rgl$& & 23 & 17 & 9 & $\lgl \pi \pi \rgl$ \\
 $\lgl \chi_+ u_\mu \rgl^2$& & 24 & 18 & & $\lgl \pi \pi \rgl$ \\
$\lgl \chi_+^3 \rgl$& & 25 & 19 & 10 & $\lgl \pi \pi \rgl$ \\
$\lgl \chi_+^2 \rgl \lgl \chi_+ \rgl$& & 26 & 20 & 11 & $\lgl \pi \pi \rgl$ \\
 $\lgl \chi_+ \rgl^3$& & 27 & 21 & & $\lgl \pi \pi \rgl$ \\
& & & & & \\
i $\lgl \chi_- \{ h_{\mu \nu},u^\mu u^\nu\} \rgl$& & 28 & 22 & 12 & $\pi
\pi \rightarrow \pi \pi$  \\
 i $\lgl \chi_-  h_{\mu \nu} \rgl \lgl u^\mu u^\nu \rgl$& & 29 & 23 & &
 $\pi \pi \rightarrow \pi \pi$ \\
 i $\lgl h_{\mu \nu} u^\mu u^\nu \rgl \lgl \chi_- \rgl$& & 30 & 24 & &
 $\pi \pi \rightarrow \pi \pi$ \\
i $\lgl h_{\mu \nu} u^\mu \chi_- u^\nu \rgl$& & 31 & 25& 13 & $\pi \pi
\rightarrow \pi \pi$ \\ 
 i $\lgl h_{\mu \nu} u^\mu \rgl \lgl \chi_- u^\nu \rgl$& & 32 & & & $\pi
 \pi \rightarrow \pi \pi$ \\ 
$\lgl u\ccdot u \chi_-^2 \rgl$& & 33 & 26 & 14 & $\pi \pi \rightarrow \pi
\pi$ \\ 
$\lgl u\ccdot u \chi_- \rgl \lgl \chi_- \rgl$& & 34 & 27 & 15 & $\pi \pi
\rightarrow \pi \pi$ \\ 
 $\lgl u\ccdot u \rgl \lgl \chi_-^2 \rgl$& & 35 & 28 & & $\pi \pi
 \rightarrow \pi \pi$\\ 
 $\lgl u\ccdot u \rgl \lgl \chi_- \rgl^2$& & 36 & & & $\pi \pi \rightarrow
 \pi \pi$\\ 
$\lgl u_\mu \chi_- u^\mu \chi_- \rgl$& & 37 & 29& 16 & $\pi \pi \rightarrow
\pi \pi$ \\ 
$\lgl u_\mu \chi_- \rgl^2$& & 38 & 30 & & $\pi \pi \rightarrow \pi \pi$\\
$\lgl \chi_-^2 \chi_+ \rgl$& & 39 & 31 & 17 & $\lgl \pi \pi \rgl$\\
$\lgl \chi_+ \rgl \lgl \chi_-^2 \rgl$& & 40 & 32 & 18 & $\lgl \pi \pi \rgl$ \\
$\lgl \chi_+ \chi_- \rgl \lgl \chi_- \rgl $ & & 41 & 33 & 19 & $\lgl \pi \pi
\rgl$ \\ 
 $\lgl \chi_+ \rgl \lgl \chi_- \rgl^2$& & 42 &  & & $\lgl \pi \pi \rgl$\\
i $\lgl \chi_- \{ \chi_{+ \, \mu},u^\mu \} \rgl$& & 43 & 34 & 20 &
$F_S^\pi (t)$ \\ 
i $\lgl \chi_- \rgl \lgl \chi_{+ \, \mu} u^\mu \rgl$& & 44 & 35
& 21 & $F_S^\pi (t)$\\
 i $\lgl \chi_{+ \mu} \rgl \lgl \chi_- u^\mu \rgl$& & 45 & 36 &
 & $F_S^\pi (t)$ \\
 $\lgl \chi_{- \mu} \rgl^2$& & 46 & 37 & & $\langle
SS \rangle $ \\
$\lgl \chi_{+ \mu}  \chi_+^\mu \rgl$& & 47 & 38 & 22 & $\langle
SS \rangle $ \\
$\lgl \chi_{+ \mu} \rgl^2$& & 48 & 39 & 23 & $\langle
SS \rangle $ \\
& & & & & \\
$\lgl (u\ccdot u)^3 \rgl$& & 49 & 40 & 24 & $ \pi \pi \rightarrow 4
\pi$ \\ 
 $\lgl (u\ccdot u)^2 \rgl \lgl u\ccdot u \rgl$& & 50 & 41 & & $ \pi \pi
 \rightarrow 4 \pi$\\ 
 $\lgl u\ccdot u \rgl^3$& & 51 & & & $ \pi \pi \rightarrow 4 \pi$\\
 $\lgl u\ccdot u u_\mu u\ccdot u u^\mu \rgl$& & 52 & 42 & & $ \pi \pi
 \rightarrow 4 \pi$\\ 
 $\lgl u\ccdot u u_\mu \rgl^2$& & 53 & 43 & & $ \pi \pi \rightarrow 4 \pi$ \\
$\lgl u\ccdot u u_\mu u_\nu u^\mu u^\nu \rgl$& & 54 & 44 &25 & $ \pi \pi
\rightarrow 4 \pi$\\ 
 $\lgl u\ccdot u u_\mu u_\nu \rgl \lgl u^\mu u^\nu \rgl$& & 55 & & & $ \pi
 \pi \rightarrow 4 \pi$\\ 
 $\lgl u\ccdot u \rgl \lgl u_\mu u_\nu \rgl^2$& & 56 & &  & $ \pi \pi
 \rightarrow 4 \pi$\\ 
 $\lgl u\ccdot u \rgl \lgl u_\mu u_\nu u^\mu u^\nu \rgl$& & 57 & 45 & & $
 \pi \pi \rightarrow 4 \pi$ \\
$\lgl u_\mu u_\nu u_\rho u^\mu u^\nu u^\rho \rgl$& & 58 & 46 & 26 & $ \pi
\pi \rightarrow 4 \pi$  \\
 $\lgl u_\mu u_\nu u_\rho \rgl^2$& & 59 & & & $ \pi \pi \rightarrow 4 \pi$ \\
 $\lgl u_\mu u_\nu u_\rho u^\mu u^\rho u^\nu \rgl$& &  60& 47 & & $ \pi \pi
 \rightarrow 4 \pi$\\ 
 $\lgl u_\mu u_\nu u_\rho \rgl \lgl u^\mu u^\rho u^\nu \rgl$& & 61 & & & $
 \pi \pi \rightarrow 4 \pi$ \\
 $\lgl u_\mu u_\nu \rgl \lgl u_\rho u^\mu u^\rho u^\nu \rgl$& & 62 &
& & $ \pi \pi \rightarrow 4 \pi$ \\
 $\lgl u_\mu u_\nu \rgl \lgl u^\mu u_\rho \rgl \lgl u^\nu u^\rho \rgl$
& &63 & & & $ \pi \pi \rightarrow 4 \pi$ \\
&& & & &\\
 i $\lgl f_{+ \mu \nu} \{u\ccdot u, u^\mu u^\nu \} \rgl$& & 64 & 48 &
 &$\gamma^* \rightarrow 4 \pi$ \\
 i $\lgl u\ccdot u \rgl \lgl f_{+ \mu \nu} u^\mu u^\nu \rgl$& & 65 &
 49 & &$\gamma^* \rightarrow 4 \pi$ \\
i $\lgl f_{+ \mu \nu} u_\rho  u^\mu u^\nu u^\rho \rgl$& & 66 & 50
 & 27 &$\gamma^* \rightarrow 4 \pi$ \\
i $\lgl f_{+ \mu \nu} u^\mu u\ccdot u u^\nu \rgl$& & 67 & 51 & 28
&$\gamma^* \rightarrow 4 \pi$ \\ 
 i $\lgl f_{+ \mu \nu} \{ u_\rho , u^\mu u^\rho u^\nu \} \rgl$& & 68
 & 52 & &$\gamma^* \rightarrow 4 \pi$ \\
 i $\lgl f_{+ \mu \nu} u_\rho \rgl \lgl  u^\mu u^\nu u^\rho \rgl$
& & 69 & & &$\gamma^* \rightarrow 4 \pi$ \\
 i $\lgl f_{+ \mu \nu} [u^\mu ,u_\rho ] \rgl \lgl u^\nu u^\rho \rgl$
& &70 & & &$\gamma^* \rightarrow 4 \pi$ \\
$\lgl u\ccdot u f_{+ \mu \nu} f_+^{\mu\nu} \rgl$& & 71 & 53 & 29 & $\gamma
\gamma \rightarrow \pi \pi$ \\ 
 $\lgl u\ccdot u \rgl \lgl f_{+ \mu \nu} f_+^{\mu\nu} \rgl$& & 72 &
 54 & & $\gamma \gamma \rightarrow \pi \pi$ \\
$\lgl f_{+ \mu \nu} u_\rho f_+^{\mu\nu} u^\rho \rgl$& & 73 & 55 & 30 &
$\gamma \gamma \rightarrow \pi \pi$ \\
 $\lgl f_{+ \mu \nu} u_\rho \rgl^2$& & 74 & & & $\gamma \gamma \rightarrow \pi \pi$ \\
$\lgl f_{+ \mu \nu} f_+^{\mu\rho} u^\nu u_\rho \rgl$& & 75 & 56 & 31 &
$\gamma \gamma \rightarrow \pi \pi$ \\
$\lgl f_{+ \mu \nu} f_+^{\mu\rho} u_\rho u^\nu \rgl$& & 76 & 57 & 32 &
$\gamma \gamma \rightarrow \pi \pi$ \\
 $\lgl f_{+ \mu \nu} f_+^{\mu\rho} \rgl \lgl u^\nu u_\rho \rgl$ & & 77 & 58
 & & $\gamma \gamma \rightarrow \pi \pi$ \\
$\lgl f_{+ \mu \nu} \left( u_\rho f_+^{\mu\rho} u^\nu + u^\nu f_+^{
    \mu \rho}  u_\rho \right) \rgl$& & 78 & 59 & 33 & $\gamma \gamma
\rightarrow \pi \pi$ \\ 
 $\lgl f_{+ \mu \nu} u_\rho \rgl \lgl f_+^{\mu\rho} u^\nu \rgl$& & 79 & & &
 $\gamma \gamma \rightarrow \pi \pi$ \\
 $\lgl f_{+ \mu \nu} u^\nu \rgl \lgl f_+^{\mu\rho} u_\rho \rgl$
& & 80 & 60 & & $\gamma \gamma \rightarrow \pi \pi$ \\
$\lgl \chi_+ f_{+ \mu \nu} f_+^{\mu\nu} \rgl$& & 81 & 61 & 34 & $\lgl VV
\rgl$ \\
 $\lgl \chi_+ \rgl \lgl f_{+ \mu \nu} f_+^{\mu\nu} \rgl$& & 82 & 62 & &
 $\lgl VV \rgl$ \\
 i $\lgl f_{+ \mu \nu} \{\chi_+, u^\mu u^\nu \} \rgl$& & 83 & 63 & & $
 F_V^\pi(t), \; K_{l3}$  \\
 i $\lgl \chi_+ \rgl \lgl f_{+ \mu \nu} u^\mu u^\nu \rgl$& & 84 & 64
 & & $ F_V^\pi(t), \; K_{l3}$ \\
i $\lgl f_{+ \mu \nu} u^\mu \chi_+ u^\nu \rgl$& & 85 & 65 & 35&  $
 F_V^\pi(t), \; K_{l3}$ \\
&& & & & \\
$\lgl f_{- \mu \nu} \left(h^{\nu\rho} u_\rho u^\mu + u^\mu u_\rho
  h^{\nu\rho} \right) \rgl$& & 86 & 66 & 36 & $K_{l4}$ \\
$\lgl f_{- \mu \nu} h^{\nu\rho} \rgl \lgl u^\mu u_\rho \rgl$& & 87
  & 67 & 37 & $K_{l4}$ \\
 $\lgl f_{- \mu \nu} u^\mu \rgl \lgl h^{\nu\rho} u_\rho \rgl$& & 88
  & 68 & & $K_{l4}$ \\
$\lgl f_{- \mu \nu} \left(u^\mu h^{\nu\rho} u_\rho +  u_\rho
  h^{\nu\rho} u^\mu \right) \rgl$& & 89 & 69 & 38 & $K_{l4}$ \\
$\lgl u\ccdot u f_{- \mu \nu} f_-^{\mu\nu} \rgl$& & 90 & 70 & 39 & $K_{l4
  \gamma} $ \\
 $\lgl u\ccdot u \rgl \lgl f_{- \mu \nu} f_-^{\mu\nu} \rgl$& & 91 &
  71 & & $K_{l4 \gamma}$ \\
$\lgl f_{- \mu \nu} u_\rho f_-^{\mu\nu} u^\rho \rgl$& & 92 & 72 & 40&
$K_{l4 \gamma}$  \\
$\lgl f_{- \mu \nu} u_\rho \rgl^2$& & 93 & & & $K_{l4 \gamma}$ \\
$\lgl f_{- \mu \nu} f_-^{\mu\rho} u^\nu u_\rho \rgl$& & 94 & 73 & 41&
$K_{l4 \gamma}$  \\
$\lgl f_{- \mu \nu} f_-^{\mu\rho} u_\rho u^\nu \rgl$& & 95 & 74 & 42&
$K_{l4 \gamma}$  \\
$\lgl f_{- \mu \nu} f_-^{\mu\rho} \rgl \lgl u^\nu u_\rho \rgl$ & & 96 & 75
& & $K_{l4 \gamma}$  \\
$\lgl f_{- \mu \nu} \left( u_\rho f_-^{\mu\rho} u^\nu + u^\nu f_-^{
    \mu \rho}  u_\rho \right) \rgl$& & 97 & 76 & 43 & $K_{l4 \gamma}$  \\
$\lgl f_{- \mu \nu} u_\rho \rgl \lgl f_-^{\mu\rho} u^\nu \rgl$
 & & 98 & & & $K_{l4 \gamma}$  \\
$\lgl f_{- \mu \nu} u^\nu \rgl \lgl f_-^{\mu\rho} u_\rho \rgl$
 & & 99 & 77 & & $K_{l4 \gamma}$  \\
i $\lgl f_{+ \mu \nu} [f_-^{ \nu \rho},h^\mu_\rho] \rgl$& & 100 &
  78 & 44& $\pi \rightarrow l\nu \gamma$ \\
i $\lgl f_{+ \mu \nu} [f_-^{ \nu \rho},f_{-\rho}^\mu] \rgl$& & 101
  & 79 & 45 & $\lgl V AA \rgl $ \\
$\lgl \chi_+ f_{- \mu \nu} f_-^{\mu\nu} \rgl$& & 102 & 80 & 46 & $\lgl A A
\rgl $ \\
 $\lgl \chi_+ \rgl \lgl f_{- \mu \nu} f_-^{\mu\nu} \rgl$& & 103 & 81
  & & $ \lgl AA \rgl $ \\
$\lgl f_{+ \mu \nu} [f_-^{\mu\nu},\chi_- ] \rgl$& & 104 & 82 & 47 & $\pi
\rightarrow l\nu \gamma$ \\
i $\lgl f_{- \mu \nu} [\chi_-, u^\mu u^\nu ] \rgl$& & 105 & 83 & 48
&$K_{l4}$ \\
 i $\lgl f_{- \mu \nu} u^\nu \rgl \lgl u^\mu \chi_- \rgl$& & 106 &
  84 & & $K_{l4}$ \\
$\lgl f_{- \mu \nu} \{ \chi_+^\mu, u^\nu \} \rgl$& & 107 &
  85 & 49 & $\lgl V AA \rgl $ \\
 $\lgl \chi_+^\mu \rgl \lgl f_{- \mu \nu} u^\nu \rgl$& & 108
  & 86 & & $\lgl V AA \rgl $ \\
$\lgl \nabla_\rho f_{- \mu \nu} \nabla^\rho f_-^{\mu\nu} \rgl$
& &109 & 87 & 50 &$\lgl AA \rgl$ \\
i $\lgl \nabla_\rho f_{+ \mu \nu} [h^{\mu\rho}, u^\nu ] \rgl$
& &110 & 88 & 51 & $ F_V^\pi (t), \; K_{l3} $ \\
i $\lgl \nabla^\mu f_{+ \mu \nu} [f_-^{ \nu \rho}, u_\rho ]
\rgl$& & 111 & 89 & 52 & $\pi \rightarrow l \nu \gamma^* $ \\
i $\lgl \nabla^\mu f_{+ \mu \nu} [h^{\nu\rho}, u_\rho ] \rgl$
& &112 & 90 & 53 & $ F_V^\pi (t), \; K_{l3} $ \\
\hline
\rule{0cm}{1.5em}\hspace{1cm} contact terms  & & & & & \\
$\lgl D_\mu \chi D^\mu \chi^\dagger \rgl$& & 113 & 91 & 54 &\\
i $\lgl F_{L \mu \nu} F_L^{ \mu \rho} F_{L \rho}^\nu \rgl +
L \rightarrow R$& & 114 & 92 & 55 & \\
$\lgl D_\rho F_{L \mu \nu} D^\rho F_L^{ \mu \nu} \rgl +
L \rightarrow R$ & & 115 & 93 & 56 &\\
\hline
\rule{0cm}{1.5em}additional contact term for $SU(3)$ & & & & & \\
$\mbox{det}(\chi)$ + h.c.& &  &94 & & \\\hline
\rule{0cm}{1.5em}additional contact term for $SU(2)$ & & & & & \\
$\lgl D_\mu \chi D^\mu \widetilde\chi \rgl$ + h.c.& &  & & 57 & \\
\end{longtable}


\begin{thebibliography}{99}
\bibitem{wein79}
S. Weinberg, {\it Physica} {\bf 96A} (1979) 327.
\bibitem{gl84}
J. Gasser and H. Leutwyler, \ap{158}{1984}{142}.
\bibitem{gl85}
J. Gasser and H. Leutwyler, \npb{250}{1985}{465}.
\bibitem{badhonnef} J. Bijnens and U.-G. Mei\ss ner,
Mini-Proceedings of the Workshop on Chiral Effective Theories,
LU-TP 99-1, \hepph{9901381}.
\bibitem{p6su2}
S. Bellucci, J. Gasser and M.E. Sainio, \npb{423}{1994}{80};\\
U. B\"urgi, \plb{377}{1996}{147}; \npb{479}{1996}{392};\\
J. Bijnens and P. Talavera, \npb{489}{1997}{387};\\
J. Bijnens, G. Colangelo, G. Ecker, J. Gasser and M.E. Sainio,
\plb{374}{1996}{210}; \npb{508}{1997}{263}; {\bf B 517} (1998) 639(E);\\
J. Bijnens, G. Colangelo and P. Talavera, \jhep{ 05}{1998}{014}.
\bibitem{p6su3} E. Golowich and J. Kambor, \npb{447}{1995}{373};
 \prd{58} {1998} {3604};\\
P. Post and K. Schilcher, \prl{79}{1997}{4088}.
\bibitem{bce3}
J. Bijnens, G. Colangelo and G. Ecker, in preparation.
\bibitem{bce4}
J. Bijnens, G. Colangelo and G. Ecker, \plb{441}{1998}{437}.
\bibitem{fs96}
H.W. Fearing and S. Scherer, \prd{53}{1996}{315}.
\bibitem{largeNc}
G. 't Hooft, \npb{772}{1974}{461}.
\bibitem{bbc}
J. Bijnens, A. Bramon and F. Cornet, \zpc{46}{1990}{499}.
\bibitem{fs95}
S. Scherer and H.W. Fearing, \prd{52}{1995}{6445}.
\bibitem{ccwz}
S. Coleman, J. Wess and B. Zumino, \pr{177}{1969}{2239};\\
C. Callan, S. Coleman, J. Wess and B. Zumino, \pr{177}{1969}{2247}.
\end{thebibliography}
\end{document}